# W+2jets production at Tevatron – VECBOS and CompHEP comparison

November 13, 1995

A. Belyaev[1], E. Boos[2], L. Dudko[3], A. Pukhov[4]

*Institute of Nuclear Physics, Moscow State University, 119899 Moscow, Russia*

**Abstract**

Results of calculation of all subprocesses in proton-antiproton collisions which contribute to the W+2jets final state are presented at Tevatron energy. The calculation has been carried out by means of the CompHEP software package. A detail comparison with VECBOS generator results for cross sections and various distributions shows an agreement at the level of Monte-Carlo accuracy. Therefore the additional independent check of VECBOS generator has been done. In complement to the VECBOS generator a new generator based on CompHEP allows to study individual subprocesses like $Wb\bar{b}$ or $Wc\bar{c}$. The last point is important, for instance, for study $Wb\bar{b}$ part of the background for single top or Standard Model Higgs signal at Tevatron.

## 1 Introduction

Reactions with jets production at Tevatron provide a very important part of physical backgrounds to a different signal processes, like strong top pair production [1, 2], electroweak bosons pair production [3, 4], electroweak single top production [5, 6, 7, 8, 9, 10, 11, 12, 13, 14, 15, 16], Standard Model Higgs production [17, 18]. In various searches VECBOS generator [19] has been used for such a background simulation. The processes with $W + 2, 3, 4$jets have been calculated in the past [20]. The calculations of the $Wb\bar{b}$ processes including b-quark mass have been carried out in [21].

In this paper we present additional detail comparison of the rates and distributions between VECBOS and CompHEP for the simplest ( from calculation point of view) case of W+2jets production taking into account all nontrivial masses. Also contributions from different parton subreactions are presented separately. CompHEP [22, 23] is known software package for automatic calculation of cross sections and distributions for particle processes in Standard Model or any model which can be easily inserted by user ( e.g. composite models, supersymmetry, model with leptoquarks).

For a comparison we did not include sea s - and c - quarks in the initial states. In fact this contribution is about 5 % from the total W+2jets rate. We also did not include any

[1]e-mail: belyaev@sgi.npi.msu.su
[2]e-mail: boos@theory.npi.msu.su
[3]e-mail: dudko@sgi.npi.msu.su
[4]e-mail: pukhov@sasha.npi.msu.su

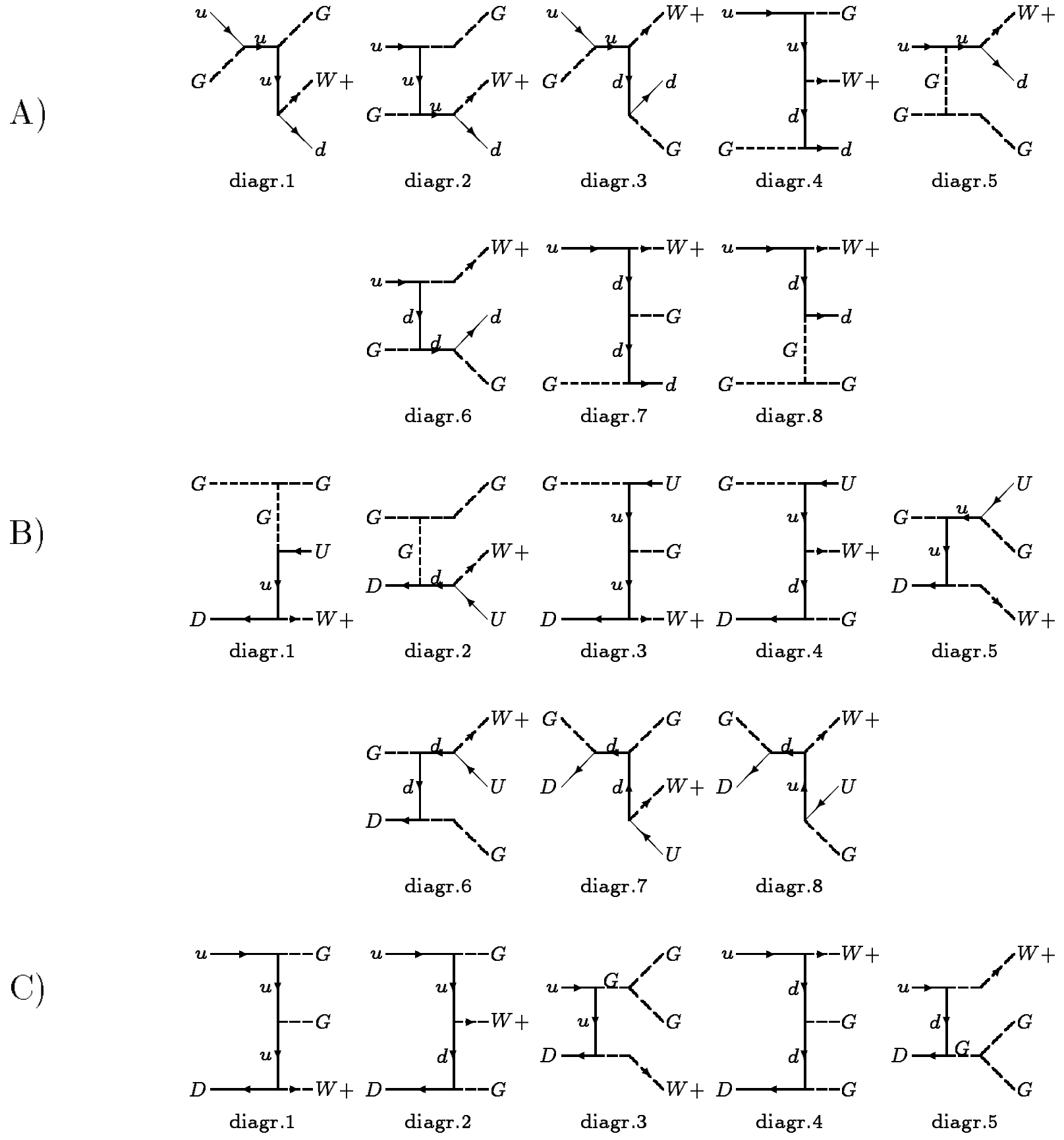

Figure 1: diagrams for the most important subreactions for W+2jets background process: A) $ug \to W^+dg$, B) $g\bar{d} \to W^+\bar{u}g$ and C) $u\bar{d} \to W^+gg$

fragmentation of partons to jets. So we considered final states just on a parton level for both VECBOS and CompHEP.

There are many electroweak diagrams in the Standard Model which contribute to the W+2jets final state. The complete set of Standard Model diagrams one can easily get using the program CompHEP. However like in VECBOS we have taken into account only the strong diagrams which provide the dominate contribution. An example of such diagrams for the most important subreactions is presented in Fig.1. In Fig.1 the CompHEP notation for particles are shown. Namely, small letters 'u' and 'd' correspond to the quark line and capital letters 'U' and 'D' correspond to the anti-quark line. Capital 'G' denotes the gluon line, 'W+' corresponds to the $W^+$-boson.

In the calculations we have used for both VECBOS and CompHEP the CTEQ2pMS set of structure functions [24, 25] with QCD scale chosen equal W-boson mass and $\lambda_{QCD} = 0.135$ GeV. For all partonic jets in the final states the $P_{Tj} > 20$ GeV cut has been used and we have used $\delta R_{jj} > 0.5$ cut for jet separation.

# 2 Rates and distributions

With assumptions mentioned above we have got the following total rate for the process $p\bar{p} \to W + 2jet$ at the 1.8 TeV Tevatron energy:

Table 1: Contributions from different subreactions for W+2jets background

| process | CompHEP $\sigma$ [pb] | CompHEP % from total rate | VECBOS % from total rate |
|---|---|---|---|
| $gg \to W^+ s\bar{c}$ | 1.097 | | |
| $gg \to W^+ d\bar{u}$ | 1.096 | | |
| $gg \to W^+ 2jet$ | 2.193 | 1.7% | 1.7% |
| $gu \to W^+ dg$ | 2.447 | | |
| $g\bar{d} \to W^+ \bar{u}g$ | 16.062 | | |
| $gq + g\bar{q} \to W^+ 2jet$ | 18.509 | 14.2% | 14.5% |
| $ug \to W^+ dg$ | 40.631 | | |
| $\bar{d}g \to W^+ \bar{u}g$ | 3.940 | | |
| $qg + \bar{q}g \to W^+ 2jet$ | 44.571 | 34.0% | 34.8% |
| $uu \to W^+ du$ | 1.323 | | |
| $\bar{d}\bar{d} \to W^+ d\bar{u}$ | 0.789 | | |
| $\bar{u}\bar{d} \to W^+ \bar{u}\bar{u}$ | 0.259 | | |
| $\bar{d}\bar{u} \to W^+ \bar{u}\bar{u}$ | 1.016 | | |
| $uu \to W^+ du$ | 1.323 | | |
| $qq + \bar{q}\bar{q} \to W^+ 2jet$ | 4.710 | 3.6% | 3.7% |
| $u\bar{d} \to W^+ b\bar{b}$ | 1.191 | | |
| $u\bar{d} \to W^+ s\bar{s}$ | 1.275 | | |
| $u\bar{d} \to W^+ c\bar{c}$ | 1.261 | | |
| $u\bar{d} \to W^+ d\bar{d}$ | 4.898 | | |
| $u\bar{d} \to W^+ u\bar{u}$ | 4.914 | | |
| $u\bar{u} \to W^+ d\bar{u}$ | 8.024 | | |
| $u\bar{u} \to W^+ s\bar{c}$ | 0.097 | | |
| $u\bar{d} \to W^+ gg$ | 33.003 | | |
| $d\bar{d} \to W^+ d\bar{u}$ | 0.099 | | |
| $d\bar{d} \to W^+ s\bar{c}$ | 0.020 | | |
| $d\bar{u} \to W^+ dd$ | 3.412 | | |
| $\bar{d}u \to W^+ b\bar{b}$ | 0.036 | | |
| $\bar{d}u \to W^+ s\bar{s}$ | 0.039 | | |
| $\bar{d}u \to W^+ c\bar{c}$ | 0.039 | | |
| $\bar{d}u \to W^+ d\bar{d}$ | 0.103 | | |
| $\bar{d}u \to W^+ u\bar{u}$ | 0.103 | | |
| $\bar{d}u \to W^+ gg$ | 0.936 | | |
| $\bar{d}\bar{d} \to W^+ \bar{d}\bar{u}$ | 1.491 | | |
| $\bar{d}\bar{d} \to W^+ s\bar{c}$ | 0.001 | | |
| $\bar{u}u \to W^+ s\bar{c}$ | 0.001 | | |
| $\bar{u}u \to W^+ \bar{d}\bar{u}$ | 0.037 | | |
| $q\bar{q} \to W^+ 2jet$ | 60.95 | 46.5% | 46.3% |

$$\text{CompHEP}: p\bar{p} \longrightarrow W + 2jet \qquad \mathbf{130.0\ pb}$$

$$\text{VECBOS}: p\bar{p} \longrightarrow W + 2jet \qquad \mathbf{127.3\ pb}$$

Therefore the difference is of order of 2% and it means the results are in an agreement within an accuracy of Monte-Carlo calculation.

Contributions from different subreactions are presented in Table 1.

Table 2: Main processes for W+2jets background

| process | cross section |
|---|---|
| $ug \rightarrow W^+dg$ | 40.6 pb ( 31.2% ) |
| $g\bar{d} \rightarrow W^+\bar{u}g$ | 16.0 pb ( 12.3% ) |
| $u\bar{d} \rightarrow W^+gg$ | 33.0 pb ( 25.4% ) |

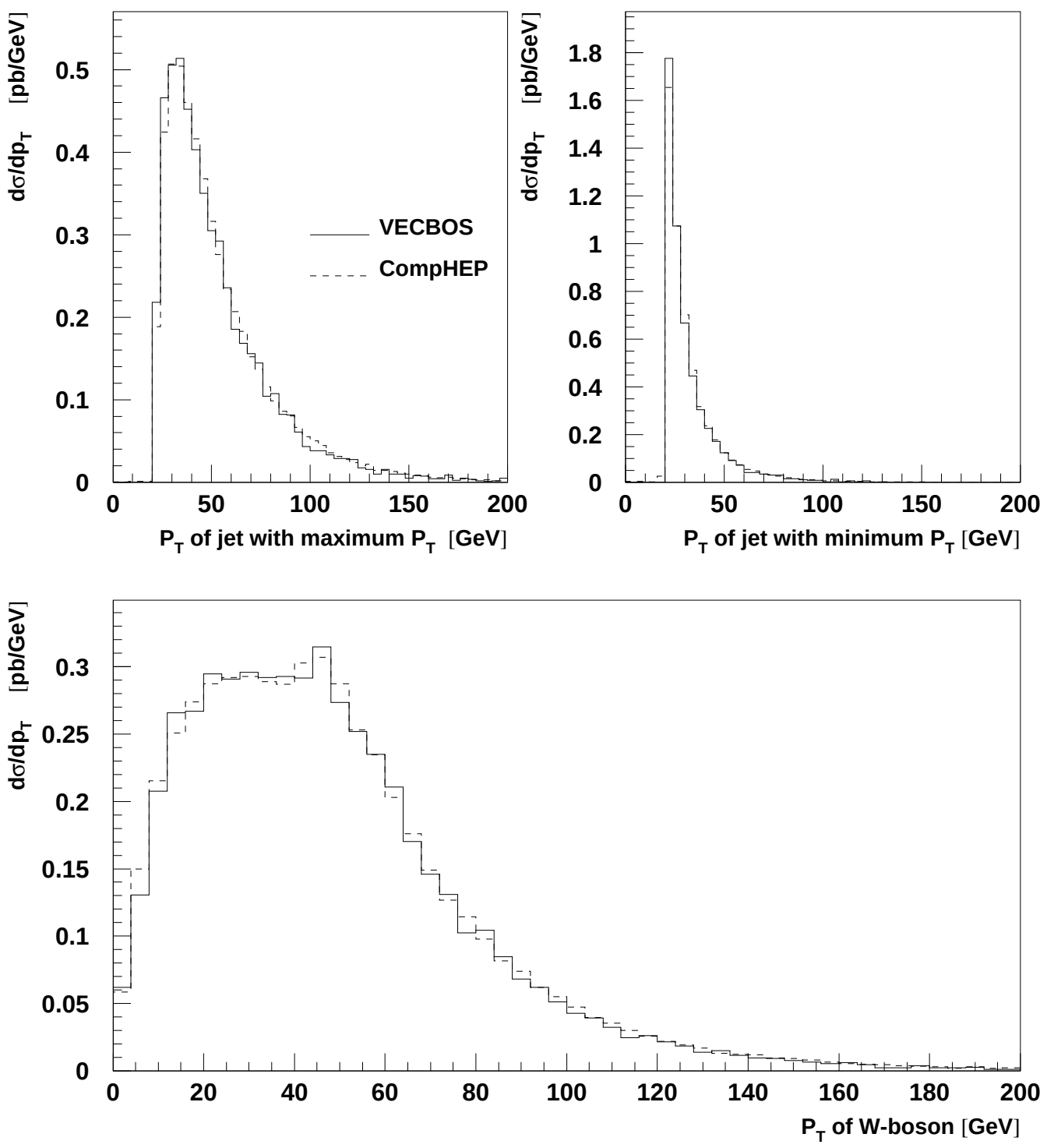

Figure 2: $P_T$ distribution of a jet with a maximum and minimum $P_T$

In all cases the integration over initial parton distributions has been performed. In the Table 1 the first initial parton is in the proton and the second one is in the antiproton. It is interesting to point out that about 68.9 % from the total rate is coming from the three dominating subprocesses (see Fig.1) listed in Table 2 once more.

In Table 1 the comparison between CompHEP and VECBOS is presented for different sets of subprocesses combined according to the initial parton states configuration. These sets are used in VECBOS. One can see a reasonable agreement for all cases.

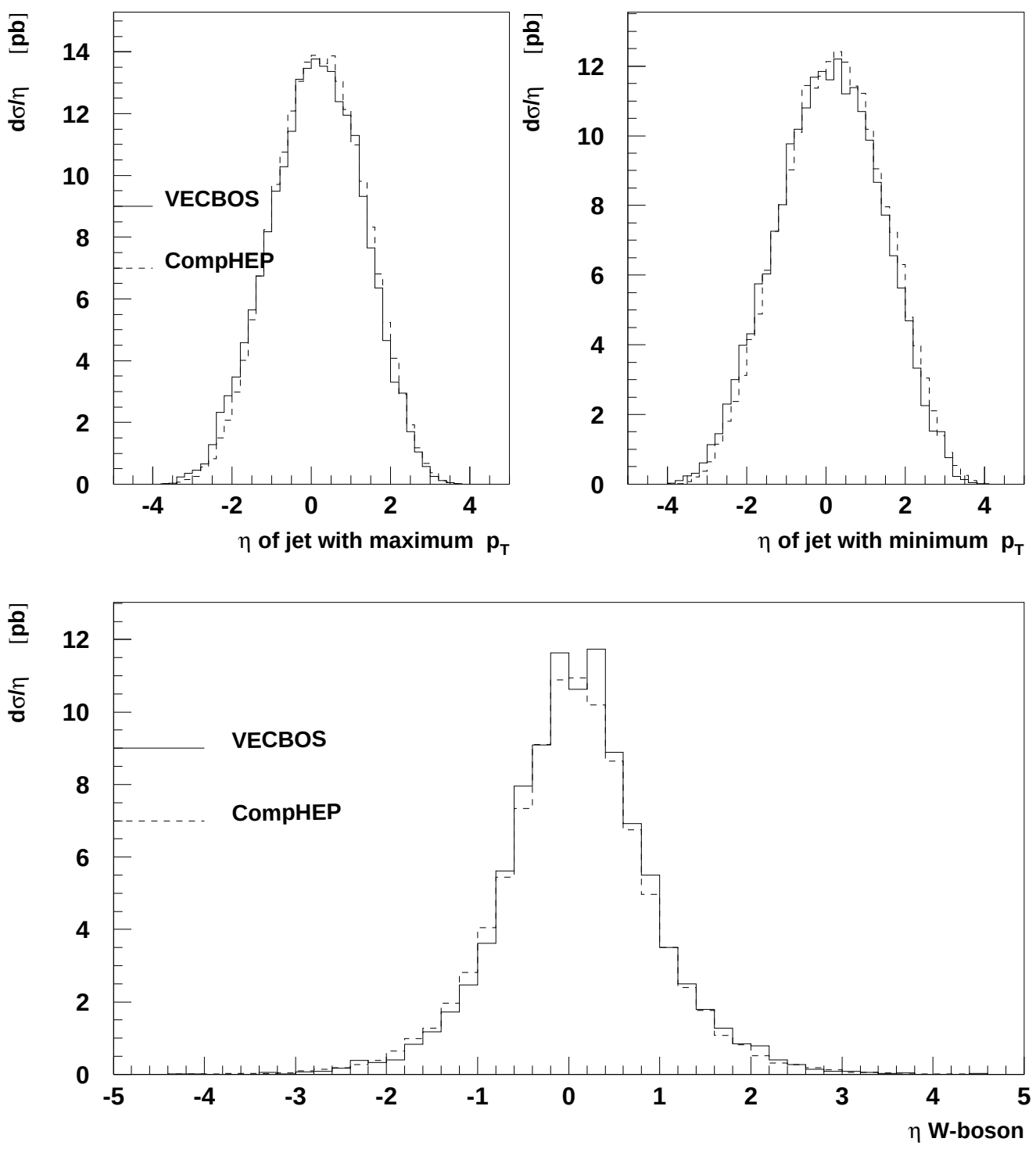

Figure 3: Pseudorapidity distribution of a jet with a maximum and minimum $P_T$

In Fig.2 the $P_T$ distribution of a jet with a maximum and minimum $P_T$ is shown. The CompHEP $P_T$ distribution is a little bit harder at high $P_T$. However this difference is of order of statistical fluctuations.

In Fig.3 pseudorapidity distributions are demonstrated. One can see that the difference between CompHEP and VECBOS for the same jets with a maximum and minimum $P_T$ is rather small like in previous case.

For various physical cases it is very important to know invariant mass distributions. The invariant mass distributions of two jets and W and jet are shown in Fig.4. One can see a very good agreement for VECBOS and CompHEP results.

# 3 Conclusions

We have calculated contributions from all subprocesses to the W+2jets production process at Tevatron using CompHEP program. For calculation of the total rates CompHEP program has been used itself while for event generation and analysis of the various distributions special event generator on the base of CompHEP package has been created.

The only QCD part from a complete set of tree level diagrams has been taken into account. For a total rate and basic distributions an agreement between CompHEP and VECBOS results at the level of statistical fluctuations has been found. It means an additional independent check of a VECBOS generator has been done. This is important because the VECBOS has been used in particular for a background simulation in the top-quark analysis and discovery [1, 2].

One can stress, however, that in the VECBOS case user can not get an information for an individual subreactions listed in the Table 1. But in some cases it can be very useful. For instance, the process with $Wb\bar{b}$ production provides an important background for searches for single top (see [16] and references therein) or SM Higgs [17, 18] in case if an effective b-tagging procedure is used.

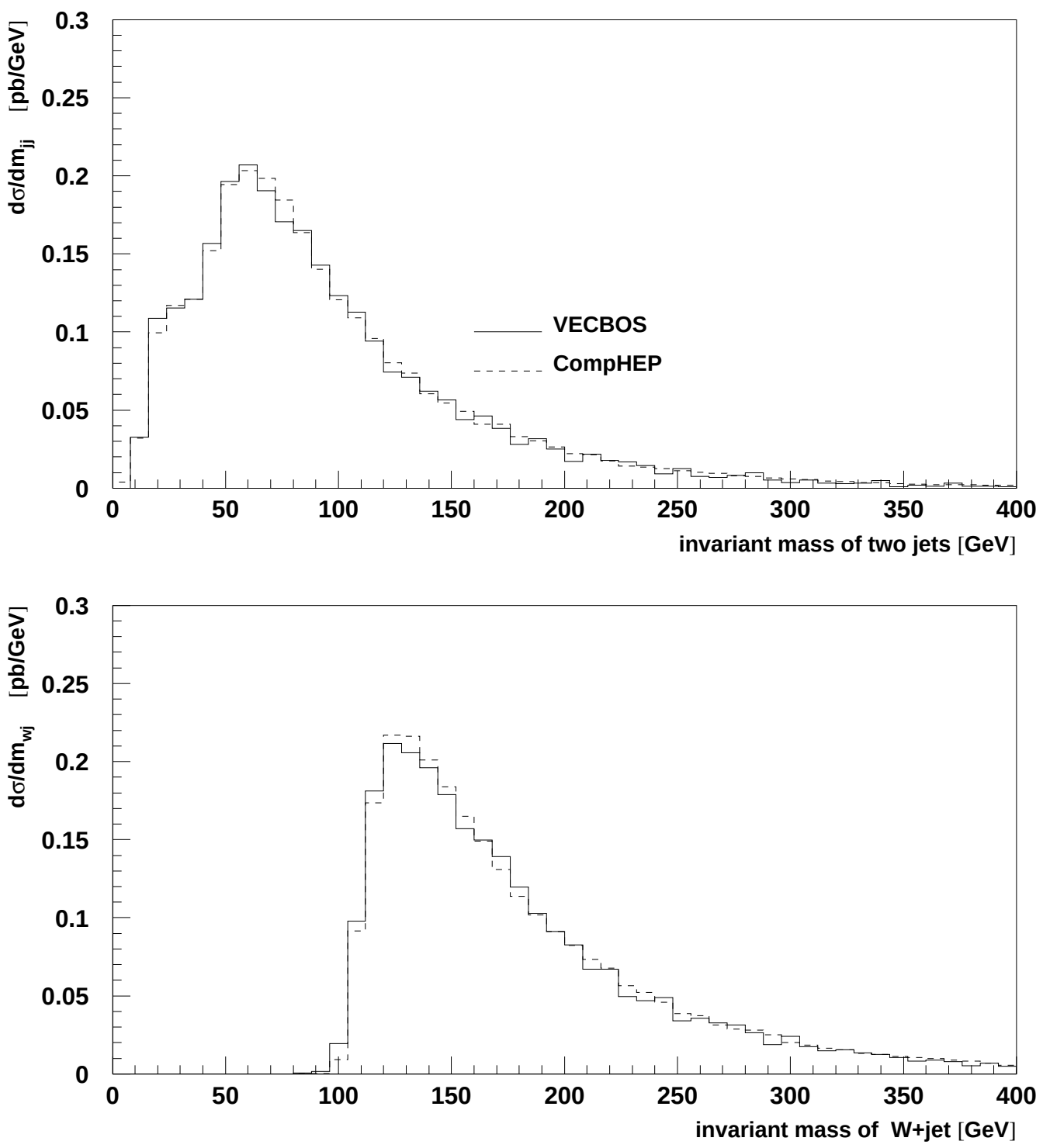

Figure 4: Invariant mass distributions of two jets and W and jet are shown

With the help of CompHEP one can separately calculate and analyse each of the subreaction, in particular, with $Wb\bar{b}$ or $Wc\bar{c}$ production taking into account fermions masses. Complete tree level calculations in Standard Model for the reactions $Wb\bar{b}$ and $Wb\bar{b}+$ jet including all nontrivial masses have been done in [26] using COmpHEP package.

# Acknowledgements

We would like to thank Pavel Ermolov, Boaz Klima, Ann Heinson and Slava Ilyin for useful discussions. We thank the DØ Collaboration for their kind hospitality during our stay at Fermilab. We acknowledge the financial support of the U.S. Department of Energy and the Ministry of Science and Technology Policy in Russia. This work has been supported in part by grants #M9B300, #a140-f from the International Science Foundation and grand and grant #93-2492 from ICFPM&INTAS.